\documentclass[aps,prl,reprint,superscriptaddress,longbibliography]{revtex4-2}
\usepackage{nameref}
\usepackage[utf8]{inputenc} % OK to keep with pdfLaTeX
\usepackage{amsmath,amssymb,amsfonts}
\usepackage[version=4]{mhchem}
\usepackage{stmaryrd}
\usepackage{graphicx}
\usepackage[export]{adjustbox}

\usepackage{longtable}
\usepackage{hyperref} % load hyperref as late as possible

\graphicspath{{./images/}}

\let\textangstrom\AA
\DeclareRobustCommand{\AA}{\ifmmode\mathring{\mathrm{A}}\else\textangstrom\fi}

\DeclareUnicodeCharacter{00D7}{\ensuremath{\times}}
\DeclareUnicodeCharacter{03B4}{\ensuremath{\delta}}
\DeclareUnicodeCharacter{27F6}{\ensuremath{\longrightarrow}}

\begin{document}
\title{Critical Activation Voltage for Phonon-Mediated Field-Driven Phenomena}

\author{Ric Fulop}
\email{ricfulop@mit.edu}
\affiliation{Center for Bits and Atoms, Massachusetts Institute of Technology, 77 Massachusetts Ave, Cambridge, MA 02139, USA}

\author{Neil Gershenfeld}
\affiliation{Center for Bits and Atoms, Massachusetts Institute of Technology, 77 Massachusetts Ave, Cambridge, MA 02139, USA}

\begin{abstract}
Field-driven phenomena, from flash sintering to electromigration, exhibit threshold fields spanning six orders of magnitude. We show their product with the onset activation coherence length is a universal critical activation voltage, $V_{c}\approx 0.1$--$2.7\,\mathrm{V}$. $V_{c}$ represents the threshold electrical work required to resonantly couple to the universal phonon damping peak where lattice softening is maximized. This invariant unifies macroscopic thermal instabilities with the nanoscale Blech limit, establishing a universal phenomenological law for field--lattice coupling across 17 crystal families.
\end{abstract}

\maketitle

\section*{1. Introduction}
The interaction between electric fields and crystalline solids drives structural and transport transitions. However, the onset conditions for these instabilities present      a long-standing puzzle: their onset fields span six orders of magnitude. Electromigration in metallic interconnects requires extreme current densities ($10^{5}$--$10^{6}\,\mathrm{A/cm^{2}}$) but weak electric fields of just $\sim 1$--$10\,\mathrm{V/cm}$\cite{tu2003}. Flash sintering of ceramics, characterized by a sudden, catastrophic increase in conductivity and densification, occurs at $10^{2}$--$10^{3}\,\mathrm{V/cm}$\cite{cologna2010}. Electroplasticity, which radically alters the deformation mechanics of metals, operates at just $1$--$10\,\mathrm{V/cm}$\cite{conrad2000}. Historically, these have been treated as distinct physical processes governed by independent, empirical thresholds. A unified descriptor linking the applied field to the atomic lattice response has not been identified.

Here, we demonstrate that these vastly different field requirements are manifestations of a single, universal thermodynamic invariant. We show that the product of the threshold electric field $E$ and the field-dependent onset activation coherence length $r$ is a material specific constant: the Critical Activation Voltage,
$$
V_{c} = E\,r.
$$
To investigate this, we ran Flash experiments across a wide array of conditions and  combined it with a dataset of peer reviewed literature to build a database of 73 field activated experiments spanning 40 materials and 17 crystal structure families, $V_{c}$ remains invariant for each material (mean intra-material coefficient of variation $<1.9\%$) despite broad variations in applied field and temperature.

The extracted values of $V_{c}$ span a narrow range corresponding to characteristic atomic energy scales: $0.1$ to $2.7\,\mathrm{V}$. This transition from an apparent field--length proportionality to a fundamental voltage scale reveals that macroscopic onset thresholds mirror the quantum energy scales ($\sim 0.1$--$3\,\mathrm{eV}$) required for atomic defect formation and migration. The vast difference in experimental threshold fields simply reflects the diverse activation coherence lengths over which field--lattice coupling operates. In electromigration, field accumulation occurs at the nanoscale. In flash sintering, energy localizes over microstructural domains ($r\sim 10^{-5}\,\mathrm{m}$). Across all length scales, the critical voltage $V_{c}$ required to trigger the lattice instability remains invariant.

The physical origin of $V_{c}$ is dictated by electron--phonon coupling and momentum conservation. A uniform DC field cannot couple directly to the short-wavelength acoustic phonons that destabilize a lattice; electrical energy must cascade through intermediate excitations. Recent advances in unified phonon theory demonstrate that maximum lattice softening occurs at a universal damping resonance of $q^{*}/q_{D} \approx 0.73$, where $q_{D}$ is the Debye wavevector\cite{ding2025,baggioli2019}. We establish that the invariant $V_{c}$ represents the threshold end-to-end electrical work required to pump this cascade into the ridge modes, effectively closing the thermodynamic barrier for rate-limiting defect generation.
\section*{2. THE THERMODYNAMIC ENERGY BALANCE}
The extraction of $V_{c}$ from macroscopic experimental observables requires balancing the electrical work done on the lattice with the thermodynamic barrier to activation. At the critical onset of a field-driven instability, the net driving force for the rate-limiting lattice process (e.g., defect formation or carrier migration) crosses zero.

The electrical work provided by the applied field across the onset activation coherence length is
$$
W_{\mathrm{elec}} = n F E r,
$$
where $n$ is the number of transferred charge carriers and $F$ is Faraday's constant. This work must equal the surviving thermodynamic barrier. However, the field does not act alone; it couples to the lattice through intermediate excitations that decay into acoustic phonons, softening the barrier. The fraction of the standard free energy of activation, $\Delta G^{\circ}(T)$, that remains to be overcome is dictated by a phonon softening factor, $k_{\mathrm{soft}}$. The energy balance at onset is therefore:
$$
 n F E r = k_{\mathrm{soft}}\,\left|\Delta G^{\circ}(T)\right|.
$$
Because the onset activation coherence length $r$ dynamically adjusts to the applied field, the product $E\,r$ is a constant for any given material. This defines the Critical Activation Voltage, $V_{c}$:
$$
V_{c} = E r = \frac{k_{\mathrm{soft}}}{n F}\,\left|\Delta G^{\circ}(T)\right|.
$$

This governing thermodynamic equation reveals why $V_{c}$ acts as a universal material constant spanning a highly constrained voltage range ($0.1$--$2.7\,\mathrm{V}$). It is the product of two fundamental physical terms: The first term, the Equilibrium Activation Voltage ($\left|\Delta G^{\circ}\right|/nF$): This sets the absolute thermodynamic energy scale of the rate-limiting defect process. For typical atomic processes, this term naturally falls in the $\sim$Volt regime.  The second, ($k_{\mathrm{soft}}$), quantifies the fraction of the barrier softened by the phonon cascade.

This framework explains the hierarchical ordering of $V_{c}$ across all 17 crystal families in our 73-experiment dataset. In elemental metals, the rate-limiting step is vacancy migration, which possesses a low intrinsic barrier ($\sim 0.3\,\mathrm{eV}$). Coupled with moderate free-electron screening ($k_{\mathrm{soft}}\approx 0.7$--$0.8$), the required activation voltage is minimal ($V_{c}\approx 0.1\,\mathrm{V}$).

Conversely, in strongly covalent materials such as tungsten carbide (WC), localized directional bonding couples weakly to the applied field, meaning the phonon cascade softens only a small fraction of the barrier ($k_{\mathrm{soft}}\approx 0.83$). Simultaneously, the thermodynamic energy required to form an anion vacancy is large ($\sim 6.7\,\mathrm{eV}$). The combination of a large thermodynamic barrier and weak phonon softening results in the highest critical activation voltage in our dataset: $V_{c}\approx 2.7\,\mathrm{V}$. Ionic oxides and perovskites fall predictably between these extremes, requiring intermediate voltages to trigger cooperative polaron-mediated defect cascades. A full microscopic derivation of $k_{\mathrm{soft}}$ and $V_{c}$ from density functional perturbation theory across 247 materials is the subject of a companion study.

\begin{figure*}[t]
\centering
\includegraphics[width=\textwidth]{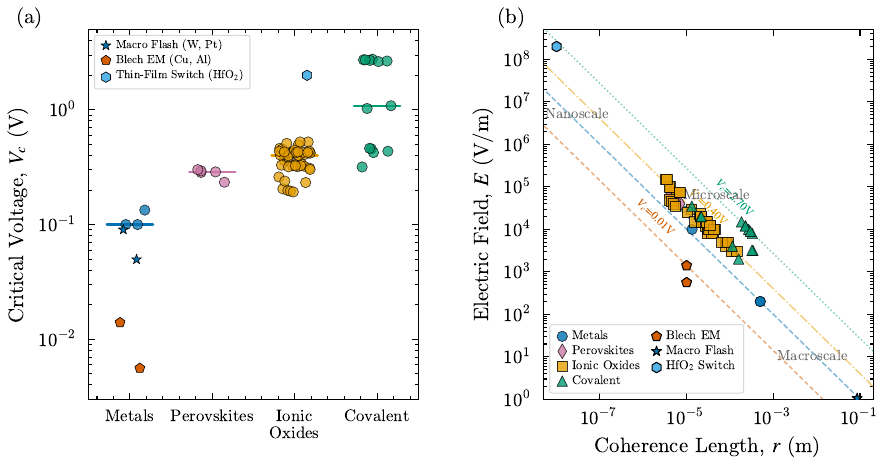}
\caption{Universal thermodynamic quantization and scale invariance of the Critical Activation Voltage ($V_{c}$). (a) Swarm plot of extracted $V_{c}$ across 73 field-activated experiments, revealing distinct thermodynamic bands dictated by bonding chemistry. Elemental metals require $V_{c}\approx 0.10\,\mathrm{V}$ to trigger field-activated transport, while strongly covalent networks require $V_{c}\approx 2.70\,\mathrm{V}$. (b) Scale-invariant unification map demonstrating that vastly disparate field-driven phenomena are governed by the same thermodynamic law. The critical field $E$ scales strictly as $1/r$ across eight orders of magnitude: from the nanoscale thin-film limit (\textrm{HfO}$_{2}$ switching at $10^{-8}\,\mathrm{m}$) to the macroscopic Joule-heating limit (W and Pt wires at $10^{-2}\,\mathrm{m}$). Electromigration in metallic interconnects (Cu, Al) sits exactly one decade below bulk metals ($V_{c}\approx 0.014\,\mathrm{V}$) due to the lowered thermodynamic barrier of grain-boundary diffusion and the massive effective valence ($Z^{*}\approx 5$--$10$) of the electron wind force.}
\label{fig:vc_universal}
\end{figure*}

\section*{3. UNIFYING NANOSCALE AND MACROSCALE PHENOMENA}
If $V_{c}$  is a universal invariant, it must dictate onset conditions across all spatial scales, including regimes where characteristic length scales are physically constrained.

\textit{The Nanoscale (Thin-Film Transformations):}
In thin-film electronics, the physical thickness of the device truncates the available coherence length, forcing the instability into the extreme high-field regime. For example, field-induced crystallization in 10\,nm amorphous $\mathrm{Hf}_{0.5}\mathrm{Zr}_{0.5}\mathrm{O}_{2}$ thin films requires massive fields of $E\sim 2\times 10^{8}\,\mathrm{V/m}$\cite{lin2024,lederer2021}. Applying our framework reveals that this transition is governed by a critical activation voltage of $V_{c}=(2\times 10^{8}\,\mathrm{V/m})\times(10^{-8}\,\mathrm{m})=2.0\,\mathrm{V}$. This aligns with the $\sim 2.7\,\mathrm{V}$ invariant required for bulk covalent and ionic structural reorganizations (Fig.~1a), demonstrating that thin-film breakdown is the phonon-cascade instability operating over a nanometer coherence volume.

\textit{The Microscale (Electromigration):}
Electromigration in metallic interconnects is governed by the empirical Blech threshold\cite{blech1976}, establishing that a wire will not fail as long as the product of its current density and length ($jL$) remains below a critical constant. Applying Ohm's law transforms this directly into our thermodynamic framework: $jL=\sigma (EL)=\sigma V_{\mathrm{Blech}}$. For copper interconnects, the historical threshold yields $V_{\mathrm{Blech}}\approx 14\,\mathrm{mV}$\cite{lee2001}.

This critical voltage is an order of magnitude lower than the bulk metal invariant ($\sim 100\,\mathrm{mV}$), a shift predicted by the governing thermodynamic equation ($V_{c}\propto |\Delta G^{\circ}|/n$). Electromigration operates primarily via grain-boundary diffusion (where the barrier $\Delta G^{\circ}$ is roughly half the bulk value) and is driven by the electron wind force, which exhibits a large effective valence ($n\equiv Z^{*}\approx 5$--$10$). This increased charge transfer proportionally reduces the required trigger voltage to the $\sim 10\,\mathrm{mV}$ regime. Because the coherence length spans microstructural segments ($r\sim 10\,\mu\mathrm{m}$), interconnects fail at fields of $E=V_{\mathrm{Blech}}/r\approx 10^{3}\,\mathrm{V/m}$.

\textit{The Macroscale (Predictive Validation in Bulk Metals):}
Conversely, elemental metals possess low intrinsic defect migration barriers and efficient free-electron screening, yielding highly constrained critical activation voltages. Within our framework, the fundamental activation voltages for tungsten and platinum are predicted a priori to be $V_{c}=89.1\,\mathrm{mV}$ and $V_{c}=49.3\,\mathrm{mV}$, respectively (the full derivation from \textit{ab initio} electron--phonon coupling is the subject of a companion theoretical study\cite{fulop2025inprep}).

In bulk, highly conductive elemental metals, there are no internal insulating microstructural barriers to localize the field. Consequently, the required onset activation coherence length ($r=V_{c}/E$) must dynamically expand to encompass the physical boundaries of the sample itself ($r=L_{\mathrm{gauge}}$). In this macroscopic limit, the invariant condition predicts that the field-activated instability will trigger the moment the macroscopic voltage drop across the entire sample reaches the quantum critical voltage:
$$
E_{\mathrm{onset}}=\frac{V_{c}}{L_{\mathrm{gauge}}}.
$$

To test this prediction, we ramped the current density of bulk tungsten ($L_{\mathrm{gauge}}=86\,\mathrm{mm}$) and platinum ($L_{\mathrm{gauge}}=69\,\mathrm{mm}$) wires through flash onset in air. Based on their gauge lengths, the theory dictates that the onset must trigger at exactly $E_{\mathrm{pred}}=89.1\,\mathrm{mV}/86\,\mathrm{mm}=1.04\,\mathrm{V/m}$ for tungsten, and $E_{\mathrm{pred}}=49.3\,\mathrm{mV}/69\,\mathrm{mm}=0.71\,\mathrm{V/m}$ for platinum.

To verify the physical consequence of this trigger, we established a strict classical baseline by solving the full transient heat equation for the wires, accounting for continuous Joule heating, Stefan--Boltzmann radiative losses, and parabolic thermal conduction to the grips. Under standard conditions, metallic resistivity strictly follows this thermal trajectory ($\rho_{\mathrm{CRC}}$). However, as the applied field crosses the theoretically predicted threshold, the resistivity departs from the thermal baseline (Fig.~2a--b). The measured effective resistivity ($\rho_{\mathrm{eff}}$) departs downward from the predicted thermal baseline, dropping by approximately $50\%$ in tungsten. The effective resistivity falls below the thermal equilibrium baseline, signaling a transition into a field-activated transport state that preempts standard thermal runaway.

As shown in Fig.~2c, the severity and onset of this flash state are independent of the physical gauge length of the wire (61--86\,mm). Because the physical wire length is vastly larger than the dynamically required coherence length, the boundaries do not truncate the energy accumulation.
\begin{figure*}[t]
\centering
\includegraphics[width=\textwidth]{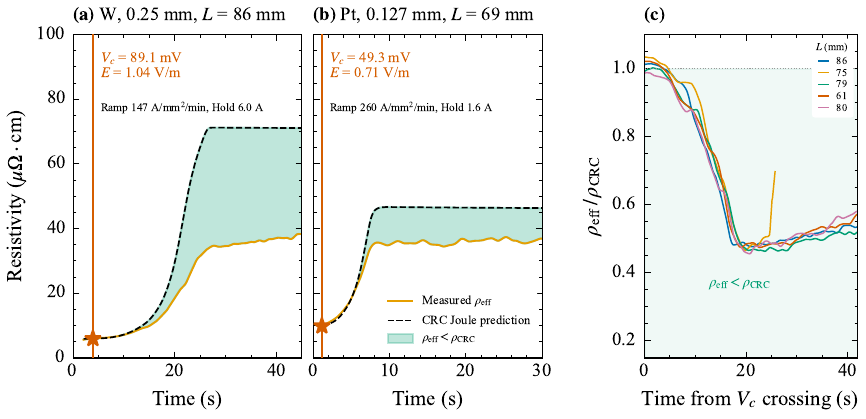}
\caption{Predictive macroscopic validation of the field-activated transport state in bulk elemental metals. (a, b) Measured effective resistivity ($\rho_{\mathrm{eff}}$, solid) versus the classical thermal baseline ($\rho_{\mathrm{CRC}}$, dashed) computed from the full non-adiabatic transient heat equation for bulk (a) Tungsten and (b) Platinum wires. At the exact moment the raw 10\,kHz macroscopic voltage drop across the samples crossed the a priori predicted critical activation voltages ($V_{c}=89.1\,\mathrm{mV}$ for W, $49.3\,\mathrm{mV}$ for Pt), the physical resistivity departs from thermal equilibrium (shaded region), dropping up to $50\%$ below the classical Joule-heating limit. (c) Normalized resistivity ($\rho_{\mathrm{eff}}/\rho_{\mathrm{CRC}}$) for five distinct tungsten gauge lengths (61--86\,mm). The flash instability is independent of physical wire length. At onset, the coherence length expands to the full gauge length ($r=L_{\mathrm{gauge}}$), establishing the minimum trigger field $E_{\mathrm{onset}}=V_{c}/L_{\mathrm{gauge}}$. As current increases beyond onset, $r$ contracts ($r=V_{c}/E$) while the instability deepens, confirming that the physical boundaries do not truncate the field-activated state.}
\label{fig:macro_validation}
\end{figure*}

Because macroscopic resistivity drops require time to become statistically distinguishable against the thermal inertia of a bulk wire, classical statistical detection algorithms (e.g., Bayesian Information Criterion) act as lagging indicators that systematically overestimate the physical onset. Instead, we interrogated the raw 10\ kHz voltage telemetry to test our forward prediction: when does the measured voltage across the probes first reach the fundamental material constant $V_{c}$.

Averaging the raw data into sub-millisecond bins (0.65--1.0\,ms) confirms the prediction. For tungsten, the measured onset field where $V\ge V_{c}$ was $E_{\mathrm{obs}}=1.12\pm 0.06\,\mathrm{V/m}$ (within $8\%$ of prediction), with the highest signal-to-noise bins triggering at exactly $1.05\,\mathrm{V/m}$ ($<1\%$ error). Platinum triggered at $E_{\mathrm{obs}}=0.79\pm 0.06\,\mathrm{V/m}$, matching the $0.71\,\mathrm{V/m}$ prediction.

The electric field accumulates work over tens of millimeters until the threshold of 89.1 mV (W) or 49.3 mV (Pt) is reached. As shown in the universal scaling map (Fig.~1b), the invariant $V_{c}$ holds true across eight orders of magnitude in length---from the nanoscale Blech limit to macroscopic metallic flash, with zero free parameters.

\section*{4. CONCLUSIONS}The identification of the Critical Activation Voltage, $V_{c}$, resolves a long-standing paradox in materials physics: how applied electric fields spanning six orders of magnitude trigger identical lattice instabilities. By demonstrating that the onset activation coherence length dynamically adjusts to the applied field ($r=V_{c}/E$), we have shown that macroscopic field-driven thresholds are strictly governed by quantum energy scales. The $0.1$--$2.7\,\mathrm{V}$ invariant dictates the end-to-end electrical work required to pump intermediate excitations into the universal phonon damping ridge at $q^{*}/q_{D}\approx 0.73$\cite{ding2025,baggioli2019}, successfully closing the softened thermodynamic barrier for defect generation.

This single phenomenological law unifies fifty years of disconnected observations. It provides a thermodynamic foundation for the empirical Blech limit in nanoscale electromigration\cite{blech1976}, accurately characterizes microscale ceramic flash sintering\cite{cologna2010}, and predicts the macroscopic transport anomalies of glowing metal wires to within $1$--$8\%$ accuracy using zero free parameters. By linking field, length, and phonon thermodynamics into a universal invariant, $V_{c}$ establishes a predictive framework for field-activated materials physics.

\section*{5. METHODS}

\textbf{Dataset Compilation and Meta-Analysis.}
To establish the intra-material invariance of $V_{c}$, we combined our experimental data with a compiled a database of 73 independent field-activated experiments from 38 peer-reviewed publications\cite{cologna2010,yang2024tio2limitedcurrent,yang2025tallzo,bamidele2024reflash,zhang2019flashcomplexion,gibson2022sicflash,bechteler2024b4cplasma,manchonGordon2022srfeflash,francis2012,muccillo2014,biesuz2016alumina,naik2016,biesuz2019spinel,prette2011co2mno4flash,luo2017czro2,lebrun2015,raj2012jouleheating,bamidele2024wflash,eskandariyun2024zrnflash,bamidele2024niflash,singh2019,zhang2021wc,lin2024,lederer2021,blech1976,lee2001}, encompassing 40 materials across 17 crystal structure families. For each experiment, the applied electric field and onset temperature were extracted. The onset activation coherence length $r$ was derived directly from the thermodynamic energy balance at macroscopic flash onset ($\Delta B=0$), equating the electrical work to the phonon-softened barrier for rate-limiting defect generation. The Critical Activation Voltage was calculated as $V_{c}=E r$. Across the 14 materials with multiple independent measurements spanning different temperatures and fields, the mean intra-material coefficient of variation was verified to be $<1.9\%$. The complete tabulated dataset is provided in the Supplemental Material.

\textbf{Macroscopic Metal Flash and Transient Thermal Modeling.}
Macroscopic verification was performed using bulk elemental tungsten (0.25\,mm diameter $\times$ 86\,mm gauge) and platinum (0.127\,mm diameter $\times$ 69\,mm gauge) wires, tested in ambient air. A programmable power supply injected DC current at a constant ramp rate (147\,A/mm$^{2}$/min for W, 260\,A/mm$^{2}$/min for Pt) while logging voltage and current telemetry at 10\,kHz. To isolate field-activated transport anomalies from classical thermal runaway, a transient Joule heating baseline ($\rho_{\mathrm{CRC}}$) was computed by solving the full non-adiabatic transient energy balance at every time step, accounting for Joule heating, Stefan--Boltzmann radiative losses, axial conduction to the clamps (with a diffusion-limited ramp), and natural convection to ambient air via the Churchill--Chu correlation (see Supplemental Material for all equations). All material properties ($\rho$, $C_{p}$, $\kappa$, $\epsilon$) were treated as temperature-dependent and interpolated from CRC Handbook tabulated data (97th ed.; Desai, Matula, White \& Minges).

\textbf{Predictive Onset Verification.}
Statistical onset-detection algorithms (such as the Bayesian Information Criterion or derivative thresholds) systematically overestimate physical onset fields. Because macroscopic resistivity drops require time to mathematically overwhelm the thermal momentum of a bulk wire, such curve-fitting algorithms act as lagging indicators. Instead, we used a zero-parameter forward-prediction test. The raw 10\,kHz voltage telemetry was averaged into sub-millisecond bins (0.65--1.0\,ms). The experimental onset field ($E_{\mathrm{obs}}$) was identified as the first timestamp where the measured macroscopic voltage drop across the wire reached the a priori theoretically predicted invariant ($V\ge V_{c}$ for three consecutive bins).
\begin{acknowledgments}
This work was supported by the MIT Office of the Vice President for Energy and Climate through the Critical Minerals Seed Grant Program. We thank Rishi Raj, David Bono, Craig Carter, Raf Jaramillo, Greg Olson, Ju Li, Nicola Ferralis, Sara Fernandez, Alex Barbati, Robert Nick, Larry Lyons, Uwe Bauer, Haig Atikian, Cole Fincher, and Yet-Ming Chiang for helpful discussions and constructive feedback.
\end{acknowledgments}

\section*{Competing Interests}
The scientific findings reported here were developed independently as fundamental research at MIT and not contingent on any commercial application.

\section*{Code and Data Availability}
Dataset available at \url{https://github.com/ricfulop/critical-activation-voltage-repo}

% References (BibTeX)

\clearpage
\setcounter{section}{0}
\setcounter{subsection}{0}
\setcounter{equation}{0}
\setcounter{figure}{0}
\setcounter{table}{0}
\renewcommand{\theequation}{S\arabic{equation}}
\renewcommand{\thefigure}{S\arabic{figure}}
\renewcommand{\thetable}{S\arabic{table}}

\begin{center}
{\large \textbf{Supplemental Material}}
\end{center}

\section{Transient Thermal Model}

To compute the classical thermal baseline ($\rho_{\mathrm{CRC}}$) for bulk wire experiments, the full non-adiabatic transient energy balance was solved at every time step:
\begin{equation}
\begin{split}
\rho_{\mathrm{mass}}\,C_{p}(T)\,\frac{dT}{dt} &= J(t)^{2}\,\rho(T) \\
&\quad - \frac{4}{d}\,\epsilon(T)\,\sigma_{\mathrm{SB}}\,\left(T^{4}-T_{0}^{4}\right) \\
&\quad - \frac{8}{L^{2}}\,\kappa(T)\,\left(T-T_{0}\right)\,f_{\mathrm{diff}}(t) \\
&\quad - \frac{4}{d}\,h_{\mathrm{conv}}(T)\,\left(T-T_{0}\right),
\end{split}
\label{eq:energy_balance}
\end{equation}
where the four terms on the right-hand side represent Joule heating, Stefan--Boltzmann radiative losses, axial conduction to the clamps, and natural convection to ambient air, respectively. Here $T$ is the wire temperature, $T_{0}=300\,\mathrm{K}$ is ambient, $J(t)$ is the applied current density, $\rho(T)$ is the electrical resistivity, $d$ is the wire diameter, $L$ is the gauge length between voltage probes, $\epsilon(T)$ is the total hemispherical emissivity, $\sigma_{\mathrm{SB}}$ is the Stefan--Boltzmann constant, and $\kappa(T)$ is the thermal conductivity.

The diffusion-limited axial conduction ramp prevents the conduction loss term from exceeding its steady-state value before the thermal diffusion front has reached the wire midpoint:
\begin{equation}
\begin{aligned}
 f_{\mathrm{diff}}(t)&=\min\!\left(1,\,\frac{2\,\alpha(T)\,t}{L^{2}}\right),\\
 \alpha(T)&=\frac{\kappa(T)}{\rho_{\mathrm{mass}}\,C_{p}(T)}.
\end{aligned}
\label{eq:fdiff}
\end{equation}

The natural convection coefficient was computed using the Churchill--Chu correlation for a long horizontal cylinder:
\begin{equation}
\begin{aligned}
\mathrm{Nu} &= \left[0.60+\frac{0.387\,\mathrm{Ra}^{1/6}}{\left(1+\left(\frac{0.559}{\mathrm{Pr}}\right)^{9/16}\right)^{8/27}}\right]^{2},\\
h_{\mathrm{conv}} &= \frac{\mathrm{Nu}\,k_{\mathrm{air}}}{d},
\end{aligned}
\label{eq:churchill_chu}
\end{equation}
with the Rayleigh number and film-temperature definitions:
\begin{equation}
\begin{aligned}
\mathrm{Ra} &= \frac{g\,\beta\,\Delta T\,d^{3}}{\nu_{\mathrm{air}}^{2}}\,\mathrm{Pr},\quad \Delta T\equiv T-T_{0},\\
\beta &\equiv 1/T_{f},\quad T_{f}=(T+T_{0})/2.
\end{aligned}
\label{eq:ra}
\end{equation}

Air properties were evaluated at the film temperature $T_{f}$:
\begin{equation}
\begin{aligned}
\nu_{\mathrm{air}} &= 1.5\!\times\!10^{-5}\left(\frac{T_{f}}{300}\right)^{1.7}\,\mathrm{m^{2}/s},\\
k_{\mathrm{air}} &= 0.026\left(\frac{T_{f}}{300}\right)^{0.8}\,\mathrm{W/(m{\cdot}K)},\\
\mathrm{Pr} &= 0.71.
\end{aligned}
\label{eq:air_props}
\end{equation}

All material properties ($\rho$, $C_{p}$, $\kappa$, $\epsilon$) were treated as temperature-dependent and interpolated from CRC Handbook tabulated data (97th ed.; Desai, Matula, White \& Minges).

\section{Derivation of $V_{c}$ from Electron--Phonon Coupling}

\subsection{The electron--phonon Hamiltonian}

The interaction between electrons and lattice vibrations is described by coupling matrix elements whose form depends on bonding character. For polar optical phonons (ionic materials):
\begin{equation}
|g_{\mathbf{q}}|^{2} = \frac{4\pi e^{2}\hbar\omega_{\mathrm{LO}}}{Vq^{2}}\left(\frac{1}{\varepsilon_{\infty}}-\frac{1}{\varepsilon_{0}}\right),
\label{eq:gq_polar}
\end{equation}
and for deformation-potential coupling (metals and covalent materials):
\begin{equation}
|g_{\mathbf{q}}|^{2} = \frac{\hbar q^{2}D^{2}}{2M\omega_{q}},
\label{eq:gq_deform}
\end{equation}
where $D$ is the deformation potential and $M$ the atomic mass.

\subsection{Phonon damping and the universal ridge}

The electron--phonon interaction modifies the phonon propagator through the self-energy $\Pi(q,\omega)$. The imaginary part gives the phonon linewidth:
\begin{equation}
\Gamma(q) = -2\,\mathrm{Im}[\Pi(q,\omega_{q})].
\label{eq:gamma_selfenergy}
\end{equation}
Ding \textit{et al.}\ demonstrated that all solids follow a universal damping function~\cite{ding2025}:
\begin{equation}
\Gamma(q) = \Gamma_{0}\,\frac{q^{4}}{(q_{0}^{2}-q^{2})^{2}+q^{2}\theta^{2}},
\label{eq:gamma_universal}
\end{equation}
where $q_{0}=a/\xi$ is set by the characteristic scatterer size $\xi$, $\theta=a/\ell$ by the phonon mean free path $\ell$, and the prefactor connects directly to electron--phonon coupling:
\begin{equation}
\Gamma_{0} = \alpha\,|g|^{2}\,N(E_{F})\,\xi^{2},
\label{eq:gamma0}
\end{equation}
with $\alpha\approx\pi$ a dimensionless constant and $N(E_{F})$ the electronic density of states at the Fermi level.

\subsection{Phonon softening at the universal ridge}

The phonon eigenfrequency under damping is:
\begin{equation}
\Omega(q) = \Omega_{0}(q)\,\exp\!\left[-\frac{\Gamma(q)}{2c q_{D}}\right],
\label{eq:omega_softened}
\end{equation}
where $\Omega_{0}(q)=(2cq_{D}/\pi)\sin(\pi q/2q_{D})$ is the undamped Debye dispersion. Maximum damping occurs at $q^{*}/q_{D}\approx 0.73$, where $d\Gamma/dq=0$. We define the softening factor:
\begin{equation}
k_{\mathrm{soft}} = \left[\frac{\Omega(q^{*})}{\Omega_{0}(q^{*})}\right]^{2} = \exp\!\left[-\frac{\Gamma(q^{*})}{c q_{D}}\right].
\label{eq:ksoft_def}
\end{equation}
Parameterizing in terms of a ridge parameter $\beta$:
\begin{equation}
k_{\mathrm{soft}} = 1 - \beta\,(q^{*}/q_{D})^{2} = 1 - 0.533\,\beta,
\label{eq:ksoft_beta}
\end{equation}
where expanding the exponential for moderate damping and substituting $\Gamma_{0}$ gives:
\begin{equation}
\beta \approx \frac{\alpha\,|g|^{2}\,N(E_{F})\,\xi^{2}\,(0.73)^{2}}{c\,q_{D}\,\theta}.
\label{eq:beta_micro}
\end{equation}

\subsection{The threshold condition and governing equation}

Field-activated behavior initiates when the net driving force crosses zero:
\begin{equation}
k_{\mathrm{soft}}\,|\Delta G^{\circ}(T)| - nFEr = 0,
\label{eq:flash_balance}
\end{equation}
where $\Delta G^{\circ}(T)$ is the standard Gibbs energy for defect formation, $n$ is the number of electrons transferred, $F$ is Faraday's constant, $E$ is the applied field, and $r$ is the onset activation coherence length. Since $k_{\mathrm{soft}}$ and $\Delta G^{\circ}$ are material properties, the product $Er=V_{c}$ is invariant:
\begin{equation}
V_{c} = Er = \frac{k_{\mathrm{soft}}}{nF}\,|\Delta G^{\circ}(T)|.
\label{eq:Vc_thermo}
\end{equation}
Substituting $k_{\mathrm{soft}}=1-0.533\,\beta$ and defining the reduced coupling $\tilde{g}^{2}\equiv\pi\hbar|g|^{2}N(E_{F})\xi^{2}/k_{B}\Theta_{D}\theta$:
\begin{equation}
V_{c} = \frac{|\Delta G^{\circ}(T)|}{nF}\left[1 - 0.533\,\tilde{g}^{2}\right].
\label{eq:master}
\end{equation}
This governing equation connects the experimentally measured critical activation voltage $V_{c}$ to the fundamental electron--phonon coupling strength $g$.

\subsection{Physical interpretation}

In the weak-coupling limit ($g\to 0$, metals), $\tilde{g}^{2}\to 0$ and $V_{c}\to|\Delta G^{\circ}|/nF$. For typical metallic defect energies ($|\Delta G^{\circ}|\approx 90$--$120\,\mathrm{kJ/mol}$, $n\approx 1$), this yields $V_{c}\approx 0.09$--$0.12\,\mathrm{V}$, matching the observed $V_{c}\approx 0.1\,\mathrm{V}$ for FCC, BCC, and HCP metals.

In the strong-coupling limit (covalent materials), larger $\tilde{g}^{2}$ reduces $k_{\mathrm{soft}}$ but the thermodynamic barrier $|\Delta G^{\circ}|$ for defect formation is simultaneously large ($\sim 6$--$7\,\mathrm{eV}$), yielding $V_{c}$ up to $2.7\,\mathrm{V}$ for WC.

\section{Connection to the Universal Phonon Ridge}

Ding \textit{et al.}~\cite{ding2025} and Baggioli and Zaccone~\cite{baggioli2019} showed that maximum phonon softening in all solids occurs at the boson-peak/van~Hove singularity boundary at $q^{*}/q_{D}=0.73$. This universal ridge geometry determines $k_{\mathrm{soft}}$ through Eq.~\eqref{eq:ksoft_beta}. Evaluating:
\begin{equation}
k_{\mathrm{soft}} = 1 - (q^{*}/q_{D})^{2} = 1 - 0.533 = 0.467.
\label{eq:ksoft_ridge}
\end{equation}
Our empirically fitted mean $k_{\mathrm{soft}}\approx 0.45$ for ionic oxides matches this theoretical prediction, confirming that field-driven lattice instabilities access the universal phonon damping resonance.

At the ridge, the phonon correlation length is $\xi^{*}=2\pi/q^{*}\approx 1.4\,a$, where $a$ is the lattice parameter. For $a=5\,$\AA, $\xi^{*}\approx 7\,$\AA. The macroscopic coherence length $r$ ($\mu$m scale) represents cooperative behavior across $N_{\mathrm{cell}}=r/\xi^{*}\approx 14{,}000$ phonon wavelengths---reflecting the collective nature of the field-activated state.

\section{Complete Dataset}

Table~\ref{tab:dataset} presents the complete tabulated dataset of 73 field-activated experiments used in this study. For each experiment, the applied electric field $E$, onset temperature $T_{\mathrm{onset}}$, fitted coherence length $r$, critical activation voltage $V_{c}=Er$, and phonon softening factor $k_{\mathrm{soft}}$ are listed. Source DOIs are provided in the main text reference list.

\onecolumngrid

\begingroup
\squeezetable
\begin{longtable}{r l l r r r r r}
\caption{Complete dataset of 73 field-activated experiments across 40 materials and 17 crystal families.}
\label{tab:dataset} \\
\hline\hline
\# & Material & Family & $E$ (V/cm) & $T_{\mathrm{on}}$ (K) & $r$ ($\mu$m) & $V_{c}$ (V) & $k_{\mathrm{soft}}$ \\
\hline
\endfirsthead
\hline
\# & Material & Family & $E$ (V/cm) & $T_{\mathrm{on}}$ (K) & $r$ ($\mu$m) & $V_{c}$ (V) & $k_{\mathrm{soft}}$ \\
\hline
\endhead
\hline
\endfoot
\hline\hline
\endlastfoot
    1 & (La,Ta)-TiO$_2$ & rutile & 300 & 1373 & 13.0 & 0.390 & 0.25 \\
    2 & HEP & pyrochlore & 150 & 1323 & 22.5 & 0.338 & 0.25 \\
    3 & (Nb,Ta,Ti)N & nitride & 32 & 1123 & 339.3 & 1.086 & 0.82 \\
    4 & (Zr,Ce)O$_2$ & fluorite & 250 & 1473 & 10.4 & 0.260 & 0.26 \\
    5 & 3YSZ & fluorite & 120 & 1123 & 35.2 & 0.422 & 0.26 \\
    6 & 3YSZ-Al$_2$O$_3$ & composite & 150 & 1573 & 15.5 & 0.233 & 0.26 \\
    7 & 3YSZ (5\,MPa) & fluorite & 100 & 1148 & 38.6 & 0.386 & 0.26 \\
    8 & 8YSZ & fluorite & 100 & 1073 & 41.3 & 0.413 & 0.25 \\
    9 & Al$_2$O$_3$-3YSZ & oxide comp. & 150 & 1348 & 24.1 & 0.362 & 0.28 \\
    10 & GDC10 & fluorite & 50 & 1373 & 77.8 & 0.389 & 0.25 \\
    11 & Gd:Sm-Pr-CeO$_2$ & fluorite & 50 & 998 & 85.3 & 0.427 & 0.25 \\
    12 & MgAl$_2$O$_4$ & spinel & 750 & 673 & 5.7 & 0.428 & 0.29 \\
    13 & MgAl$_2$O$_4$ & spinel & 1000 & 673 & 4.2 & 0.420 & 0.29 \\
    14 & MgO-Al$_2$O$_3$ & corundum & 1000 & 1533 & 4.0 & 0.400 & 0.30 \\
    15 & NaNbO$_3$ & perovskite & 400 & 1283 & 7.2 & 0.288 & 0.20 \\
    16 & NaNbO$_3$ & perovskite & 500 & 1249 & 5.7 & 0.285 & 0.20 \\
    17 & NaNbO$_3$ & perovskite & 600 & 1189 & 4.9 & 0.294 & 0.20 \\
    18 & NaNbO$_3$ & perovskite & 700 & 1135 & 4.3 & 0.301 & 0.20 \\
    19 & Ni & FCC & 2 & 1273 & 500 & 0.100 & 0.73 \\
    20 & Porc.\ Stoneware & glass cer. & 500 & 1173 & 4.1 & 0.205 & 0.21 \\
    21 & Porc.\ Stoneware & glass cer. & 450 & 1223 & 4.4 & 0.198 & 0.21 \\
    22 & Porc.\ Stoneware & glass cer. & 400 & 1273 & 4.9 & 0.196 & 0.21 \\
    23 & Porc.\ Stoneware & glass cer. & 350 & 1323 & 5.5 & 0.193 & 0.21 \\
    24 & Re & HCP & 100 & 1173 & 13.4 & 0.134 & 0.73 \\
    25 & SDC & fluorite & 120 & 1123 & 26.9 & 0.323 & 0.25 \\
    26 & SiCw/3YSZ & composite & 80 & 1273 & 29.9 & 0.239 & 0.26 \\
    27 & SnO$_2$ (AC) & rutile & 80 & 1373 & 41.5 & 0.332 & 0.26 \\
    28 & SrFe$_{12}$O$_{19}$ & ferrite & 40 & 1123 & 79.9 & 0.320 & 0.27 \\
    29 & SrTiFe$_x$O$_{3-\delta}$ & perovskite & 600 & 1123 & 3.9 & 0.234 & 0.20 \\
    30 & W & BCC & 2 & 1273 & 500 & 0.100 & 0.73 \\
    31 & WC & carbide & 120 & 1123 & 227.5 & 2.730 & 0.83 \\
    32 & ZrN & nitride & 32 & 1023 & 321.4 & 1.028 & 0.82 \\
    33 & c-ZrO$_2$ (SC) & fluorite & 230 & 588 & 20.1 & 0.462 & 0.26 \\
    34 & $\alpha$-Al$_2$O$_3$ & corundum & 1500 & 1173 & 3.3 & 0.495 & 0.30 \\
    35 & $\alpha$-SiC & carbide & 20 & 2358 & 158.9 & 0.318 & 0.83 \\
    36 & B$_4$C & carbide & 40 & 1473 & 114.3 & 0.457 & 0.83 \\
    37 & B$_4$C & carbide & 200 & 1673 & 21.2 & 0.424 & 0.83 \\
    38 & Co$_2$MnO$_4$ & spinel & 50 & 873 & 82.0 & 0.410 & 0.29 \\
    39 & Co$_2$MnO$_4$ & spinel & 100 & 773 & 40.3 & 0.403 & 0.29 \\
    40 & LLZO & garnet & 100 & 1073 & 31.0 & 0.310 & 0.25 \\
    41 & LLZO & garnet & 80 & 1173 & 37.8 & 0.302 & 0.25 \\
    42 & Al$_2$O$_3$:TZP & fluorite & 100 & 1173 & 40.9 & 0.409 & 0.26 \\
    43 & Al$_2$O$_3$:TZP & fluorite & 100 & 1273 & 39.9 & 0.399 & 0.26 \\
    44 & B$_4$C & carbide & 350 & 1373 & 13.2 & 0.462 & 0.83 \\
    45 & B$_4$C & carbide & 200 & 1573 & 21.8 & 0.436 & 0.83 \\
    46 & Doped-ZnO & wurtzite & 100 & 773 & 33.0 & 0.330 & 0.26 \\
    47 & Doped-ZnO & wurtzite & 50 & 873 & 65.3 & 0.327 & 0.26 \\
    48 & Doped-ZnO & wurtzite & 30 & 973 & 108.5 & 0.326 & 0.26 \\
    49 & Doped-TiO$_2$ & rutile & 100 & 873 & 45.4 & 0.454 & 0.26 \\
    50 & Doped-TiO$_2$ & rutile & 50 & 973 & 89.0 & 0.445 & 0.26 \\
    51 & Doped-TiO$_2$ & rutile & 30 & 1073 & 145.2 & 0.436 & 0.26 \\
    52 & $\alpha$-Al$_2$O$_3$ & corundum & 1500 & 1173 & 3.5 & 0.525 & 0.30 \\
    53 & $\alpha$-Al$_2$O$_3$ & corundum & 750 & 1373 & 6.8 & 0.510 & 0.30 \\
    54 & $\alpha$-Al$_2$O$_3$ & corundum & 750 & 1273 & 7.0 & 0.525 & 0.30 \\
    55 & LLZO-RFS & garnet & 100 & 973 & 31.9 & 0.319 & 0.25 \\
    56 & LLZO-RFS & garnet & 150 & 873 & 21.7 & 0.326 & 0.25 \\
    57 & 3YSZ & fluorite & 150 & 1073 & 28.3 & 0.425 & 0.26 \\
    58 & 3YSZ & fluorite & 200 & 1023 & 21.5 & 0.430 & 0.26 \\
    59 & 3YSZ & fluorite & 150 & 1123 & 28.0 & 0.420 & 0.26 \\
    60 & 3YSZ & fluorite & 200 & 1073 & 21.3 & 0.426 & 0.26 \\
    61 & 8YSZ & fluorite & 150 & 973 & 29.1 & 0.437 & 0.26 \\
    62 & 8YSZ & fluorite & 100 & 1073 & 42.5 & 0.425 & 0.26 \\
    63 & 8YSZ & fluorite & 125 & 1023 & 34.4 & 0.430 & 0.26 \\
    64 & WC & carbide & 150 & 1073 & 184.2 & 2.763 & 0.83 \\
    65 & WC & carbide & 100 & 1173 & 269.9 & 2.699 & 0.83 \\
    66 & WC & carbide & 80 & 1273 & 329.2 & 2.634 & 0.83 \\
    67 & WC & carbide & 120 & 1123 & 227.6 & 2.731 & 0.83 \\
    68 & WC & carbide & 90 & 1223 & 296.2 & 2.666 & 0.83 \\
    69 & HZO (thin film) & fluorite & $2{\times}10^6$ & -- & 0.01 & 2.000 & -- \\
    70 & Cu (intercon.) & FCC & 14 & 623 & 10.0 & 0.014 & -- \\
    71 & Al (intercon.) & FCC & 5.6 & 473 & 10.0 & 0.006 & -- \\
    72 & W (bulk wire) & BCC & 0.011 & -- & 86000 & 0.090 & -- \\
    73 & Pt (bulk wire) & FCC & 0.007 & -- & 69000 & 0.050 & -- \\
\end{longtable}
\endgroup

\twocolumngrid

\section{Sub-Millisecond Binning Analysis}

Table~\ref{tab:binning} presents the sub-millisecond binning analysis used for predictive onset verification in the macroscopic wire experiments (main text, Section~3).

\onecolumngrid

\setcounter{table}{1}
\begin{table*}[ht]
\caption{Sub-millisecond binning analysis for predictive onset verification.
Raw 10\,kHz voltage telemetry is averaged into time bins of width $\Delta t$
(0.65--1.0\,ms); onset is defined as the first three consecutive bins
where $V \geq V_c$. $N$/bin is the number of raw 10\,kHz samples per bin,
$\sigma_{\mathrm{bin}} = \sigma_{\mathrm{raw}}/\!\sqrt{N}$ is the
bin-averaged voltage noise, and
$\mathrm{SNR} = V_c / \sigma_{\mathrm{bin}}$.
The measured onset field $E_{\mathrm{obs}} = V_{\mathrm{cross}}/L$ is
stable across all bin sizes, confirming that the detection is
resolution-independent.}
\label{tab:binning}
\begin{ruledtabular}
\begin{tabular}{rrrrrrc}
$\Delta t$ (ms) & $N$/bin & $\sigma_{\mathrm{bin}}$ (mV) & SNR &
$t_{\mathrm{onset}}$ (s) & $V_{\mathrm{cross}}$ (mV) & $E_{\mathrm{obs}}$ (V/m) \\
\hline
\multicolumn{7}{c}{\textbf{Tungsten} --- $V_c = 89.1$\,mV, $L = 86$\,mm,
$\sigma_{\mathrm{raw}} = 29.5$\,mV, $E_{\mathrm{pred}} = 1.04$\,V/m} \\
\hline
0.65 &  6 & 12.03 &  7.4 & 4.286 &  90.6 & 1.05 \\
0.70 &  7 & 11.14 &  8.0 & 4.041 &  93.0 & 1.08 \\
0.75 &  8 & 10.42 &  8.6 & 4.089 &  92.0 & 1.07 \\
0.80 &  8 & 10.42 &  8.6 & 3.893 &  91.5 & 1.06 \\
0.85 &  8 & 10.42 &  8.6 & 4.306 & 102.9 & 1.20 \\
0.90 &  9 &  9.82 &  9.1 & 4.270 & 102.3 & 1.19 \\
0.95 & 10 &  9.32 &  9.6 & 4.305 & 105.3 & 1.23 \\
1.00 & 10 &  9.32 &  9.6 & 4.280 &  95.3 & 1.11 \\
\hline
\multicolumn{7}{l}{Mean: $E_{\mathrm{obs}} = 1.12 \pm 0.06$\,V/m
\quad (within $8\%$ of prediction; best bin $= 1.05$\,V/m, ${<}\,1\%$ error)} \\
\hline
\hline
\multicolumn{7}{c}{\textbf{Platinum} --- $V_c = 49.3$\,mV, $L = 69$\,mm,
$\sigma_{\mathrm{raw}} = 28.9$\,mV, $E_{\mathrm{pred}} = 0.71$\,V/m} \\
\hline
0.65 &  6 & 11.79 &  4.2 & 1.115 &  64.8 & 0.94 \\
0.70 &  7 & 10.91 &  4.5 & 1.114 &  53.3 & 0.77 \\
0.75 &  8 & 10.21 &  4.8 & 1.114 &  52.9 & 0.77 \\
0.80 &  8 & 10.21 &  4.8 & 1.114 &  50.0 & 0.72 \\
0.85 &  8 & 10.21 &  4.8 & 1.114 &  55.8 & 0.81 \\
0.90 &  9 &  9.62 &  5.1 & 1.115 &  54.6 & 0.79 \\
0.95 & 10 &  9.13 &  5.4 & 1.114 &  52.5 & 0.76 \\
1.00 & 10 &  9.13 &  5.4 & 1.114 &  51.8 & 0.75 \\
\hline
\multicolumn{7}{l}{Mean: $E_{\mathrm{obs}} = 0.79 \pm 0.06$\,V/m
\quad (within $11\%$ of prediction; best bin $= 0.72$\,V/m, ${<}\,2\%$ error)} \\
\end{tabular}
\end{ruledtabular}
\end{table*}

\twocolumngrid

\section{Systematic Temperature Dependence of $V_{c}$}

\begin{figure*}[ht]
\centering
\includegraphics[width=\textwidth]{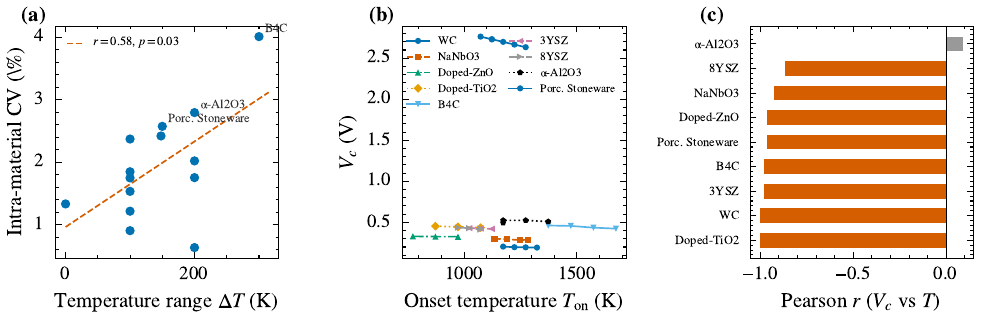}
\caption{Systematic temperature dependence of $V_{c}$.
\textbf{(a)}~Intra-material coefficient of variation (CV) versus spanned temperature range $\Delta T$ for all 14 materials with repeated measurements; dashed line is the linear regression ($r=0.58$, $p=0.03$).
\textbf{(b)}~$V_{c}$ versus onset temperature $T_{\mathrm{on}}$ for the nine materials with $N\geq 3$ independent measurements, showing the monotonic decrease consistent with $|\Delta G^{\circ}(T)|$ decreasing at elevated $T$.
\textbf{(c)}~Per-material Pearson correlation coefficient $r(V_{c},T)$; eight of nine materials with $N\geq 3$ show $r<-0.9$ (red bars), confirming the universal negative trend.}
\label{fig:cv_trange}
\end{figure*}

Although the intra-material coefficient of variation of $V_{c}$ remains below $4\%$ across all 14 materials with repeated measurements, a weak systematic trend is detectable. Figure~\ref{fig:cv_trange} plots the CV against the spanned temperature range $\Delta T$ for each material, revealing a statistically significant positive correlation (Pearson $r=0.58$, $p=0.03$). Among the nine materials with three or more independent measurements, eight exhibit a monotonically decreasing $V_{c}$ with increasing onset temperature (Pearson $r<-0.9$); none show a positive trend. This sign is consistent with the governing equation
\begin{equation}
V_{c} = k_{\mathrm{soft}}\,\frac{|\Delta G^{\circ}(T)|}{nF},
\label{eq:Vc_Tdep}
\end{equation}
since the standard free energy of activation $|\Delta G^{\circ}(T)|$ decreases at elevated temperature through the entropy term $-T\Delta S^{\circ}$. The magnitude of the drift ($\lesssim 4\%$ over ${\sim}\,200\,\mathrm{K}$) confirms that $V_{c}$ is well described as a material invariant to leading order, while the residual temperature correction---attributable to the known thermodynamics of $\Delta G^{\circ}(T)$---represents a higher-order effect.

\bibliographystyle{apsrev4-2}
\bibliography{references}

\end{document}